\documentstyle[12pt]{article}

\newcommand{\be}{\begin{equation}}
\newcommand{\ee}{\end{equation}}
\newcommand{\bea}{\begin{eqnarray}}
\newcommand{\eea}{\end{eqnarray}}
\newcommand{\vs}[1]{\vspace{#1 mm}}

\renewcommand{\a}{\alpha}
\renewcommand{\b}{\beta}

\def\bbox{{\,\lower0.9pt\vbox{\hrule \hbox{\vrule height 0.2 cm
\hskip 0.2 cm \vrule height 0.2 cm}\hrule}\,}}
\newcommand{\dsl}{\pa \kern-0.5em /}

\newcommand{\pa}{\partial}

\newcommand{\nn}{\nonumber\\}

\def\tr{{\rm tr}}

\begin{document}

\topmargin 0pt
\oddsidemargin 5mm

\renewcommand{\thefootnote}{\fnsymbol{footnote}}
\begin{titlepage}

\setcounter{page}{0}
\begin{flushright}
ECM-UB-97/35 \\
QMW-97-37\\
hep-th/9711205
\end{flushright}

\vs{5}
\begin{center}
{\Large BPS BOUNDS FOR WORLDVOLUME BRANES}
\vs{10}

{\large
Jerome P. Gauntlett
} \\
\vs{5}
{\em Department of Physics, Queen Mary and Westfield College, \\
University of London, Mile End Road, \\
London E1 4NS, UK} \\
\vs{7}
and \\
\vs{7}
{\large
Joaquim Gomis and Paul K. Townsend\footnote{On
leave from DAMTP,  University of Cambridge, U.K.}
} \\
\vs{5}
{\em Departament ECM, Facultat de F{\'\i}sica, \\
Universitat de Barcelona and Institut de F{\'\i}sica d'Altes Energies, \\
Diagonal 047, E-08028 Barcelona, Spain }\\ 
\end{center}
\vs{10}
\centerline{{\bf{Abstract}}}

The worldvolume field equations of M-branes and D-branes are known to 
admit p-brane soliton solutions. These solitons are shown to saturate a
BPS-type bound on their p-volume tensions, which are expressed in terms
of central charges that are expected to appear in the worldvolume
supertranslation algebra. The cases we consider include vortices, `BIons',
instantons and dyons (both abelian and non-abelian), and the string
boundaries of M-2-branes in the M-5-brane.

\end{titlepage}
\newpage
\renewcommand{\thefootnote}{\arabic{footnote}}
\setcounter{footnote}{0} 

\section{Introduction}

In the past few years many supersymmetric field theories have been reformulated
as worldvolume field theories on branes or on their intersections. One
aspect of this reformulation of a supersymmetric field theory is that its
`BPS-saturated' states, some of which correspond to 1/2 supersymmetric classical
solutions, acquire a spacetime interpretation as intersections with other
branes. The prototype is an electric charge on a Type II D-brane, which acquires
an interpretation as the endpoint of a `fundamental' Type II string 
\cite{polch}. This has its M-theory analogue in the interpretation of a
self-dual string in the M-5-brane as the boundary of an M-2-brane
\cite{Strom,PKT}. Other examples are vortices on M-2-branes or M-5-branes, which
acquire an interpretation as 0-brane or 3-brane intersections with a second
M-2-brane or M-5-brane \cite{paptown}. In all these cases, the 1/2
supersymmetric solutions of the worldvolume field theory of a single D-brane or
M-brane have now been found \cite{CM,HLW1,GWG,HLW2}.

A remarkable feature of these `worldvolume solitons' is that, while ostensibly
just solutions of some (p+1)-dimensional field theory, they in fact suggest
their own 10 or 11 dimensional {\sl spacetime} interpretation. This arises
from the fact that the world-volume scalars determine the spacetime embedding. In
a recent paper it was pointed out that the spacetime interpretation is already
implicit in the central charge structure of the worldvolume supersymmetry
algebra \cite{BGT}. An example is the D=6 (2,0) worldvolume supersymmetry
algebra of the M-5-brane. Allowing for all possible p-brane charges we have
\cite{HLW2}
\be\label{introa}
\{Q_\a^I,Q_\b^J\} = \Omega^{IJ} P_{[\a\b]} + Y^{[IJ]}_{[\a\b]} +
Z^{(IJ)}_{(\a\b)}, 
\ee
where $\a,\b =1,\dots,4$ is an index of $SU^*(4)\cong Spin(5,1)$ and
$I=1,\dots,4$ is an index of $Sp(2)$, with $\Omega^{IJ}$ being its
invariant antisymmetric tensor. The $Y$-charge, satisfying 
$\Omega_{IJ}Y^{[IJ]}=0$,
is a worldvolume 1-form carried by worldvolume strings and the $Z$-charge is a
worldvolume self-dual 3-form carried by worldvolume 3-branes. The
representations of the $R$-symmetry group $Sp(2)$ encode the
possible interpretations of the solitons carrying these charges as
intersections of other objects with the M-5-brane \cite{BGT}. For example, the
string charge, being a 5-vector of $Spin(5)$, can be viewed as a 1-form
in the 5-space transverse to the M-5-brane worldvolume in spacetime. It
therefore defines a direction in this space which may be identified as the
direction in which an M-2-brane `leaves' the M-5-brane, consistent with the
interpretation of the worldvolume string as an M-2-brane boundary. Similarly, 
the 3-brane charge can be viewed as a transverse 2-form, consistent with its
spacetime interpretation as the intersection with another M-5-brane.

The results we report on here arose from a consideration of the way in which the
magnitudes of the p-form charges carried by worldvolume solitons are expressed
as integrals of charge densities constructed from worldvolume fields. As we
shall argue, the correct expressions for the magnitudes of the charges
carried by the string and 3-brane in the M-5-brane are
\be\label{charges}
Y = \int \! dX\wedge H \qquad 
Z = {i\over2}\int \! dU \wedge d\bar U
\ee
where $H=dA$ is the worldvolume three form 
 field strength and $X$ and $U$ are, respectively,
real and complex worldvolume scalar fields describing fluctuations transverse to
the M-5-brane worldvolume in spacetime. The integrals are over the subspaces
of the 5-dimensional worldspace transverse to the p-brane
solitons. Note that $Y$ is not given simply by an integral of $H$ over a 
3-sphere surrounding the string in the M-5-brane, as one might naively have
expected. It also includes a dependence on one scalar field, as required by its
identification with the magnitude of a 1-form in the space transverse to the
M-5-brane\footnote{The string charge $Y$ also appears in \cite{DVV}, 
but in the context of the linearised five-brane theory in the
light-cone gauge. Note also that the strings considered there
are non-self-dual dipole strings with vanishing $Y$ charge.}.
Similarly, the dependence of the 3-brane charge $Z$ on a complex
scalar is required by its interpretation as a transverse 2-form. 

To establish the correctness of the above expressions, and similar expressions
for the magnitudes of charges carried by p-branes in other M-theory and Type II
branes, one could explicitly construct the supersymmetry generators as Noether
charges and determine their algebra directly\footnote {For M-branes, 
the result of such a calculation is
in principle implied
 by the results of \cite{azc,sorokin}.
 One should consider a  combination  of p-form charges 
associated to a $\frac 14$ supersymmetric configuration of 
intersecting M-branes
and then project onto the 
subspace spanned by the supercharges linearly realized on one brane.}.
Here we shall take an alternative 
path by showing that the p-volume tensions of worldvolume p-brane solitons are
bounded from below by expressions that are precisely of the form
(\ref{charges}). The method is similar to that employed in \cite{Bog} but
differs in essential respects owing to the fact that brane actions are generally
non-quadratic in derivatives. However, the configurations saturating the bound
are precisely those satisfying first-order BPS-type equations, the solutions of
which are (in this context) the worldvolume solitons found in
\cite{CM,HLW1,GWG,HLW2} which are known to preserve 1/2 of the
supersymmetry\footnote{The solution in \cite{CM,GWG} of relevance here was shown
to be supersymmetric in \cite{HLW1,peet}.}. 

In the case of a single D-brane the worldvolume hamiltonian involves the
Born-Infeld (BI) $U(1)$ 2-form field strength $F$. The particles on the
D-brane that carry the electric $U(1)$ charges were called `BIons' in
\cite{GWG}, and we shall adopt this terminology here for the supersymmetric
solutions. BIons satisfy a
BPS-type equation involving $F$, which we shall call the `abelian BIon'
equation, but there is
 a natural non-abelian extension 
 as appropriate to multiple coincident D-branes. There is  
also a natural non-abelian extension of the D-brane hamiltonian, natural in the
sense that the energy bound of the abelian case continues to hold but is
saturated by solutions of the non-abelian BIon equations. 
One can simply take this as the definition of the non-abelian BI theory in each
case of interest. One then finds, for example, that the
monopole and dyon solutions of (3+1)-dimensional supersymmetric non-abelian
gauge theories are also solutions of the non-abelian D-3-brane worldvolume
equations. This approach is physically compelling and is 
simpler than previous investigations of
the effects of BI `corrections' on BPS monopoles \cite{Nakamura}, 
but it remains
to be seen whether our definition of the non-abelian BI theory 
accords with other definitions e.g. \cite{Tseytlin}.  

The worldvolume field 
equations of M-theory or Type II branes depend on the M-theory 
or Type II background. The worldvolume solitons found in \cite{CM,HLW1,GWG,HLW2}
are 1/2 supersymmetric solutions of the worldvolume  field
equations in an M-theory or
Type II spacetime vacuum, by which we mean D=11 or D=10 Minkowski space with all
other 
space-time fields vanishing. Since an M-theory or 
Type II brane preserves half the
supersymmetry of the spacetime vacuum, its worldvolume supersymmetry algebra has
16 supercharges\footnote{The 16 non-linearly realised supersymmetries will not
play a direct role in this paper.}. Each 1/2 supersymmetric worldvolume solitons
must therefore correspond to some  p-form in one of the supersymmetry algebras
with 16 supersymmetries considered in \cite{BGT}. The converse is not true,
however, since some charges correspond to brane boundaries, which are normally
determined by imposing boundary conditions rather than by solving field
equations. An interesting exception to this is provided by the endpoints of
multiple coincident D-strings on  D-3-branes, which are determined by a solution
of Nahm's equations \cite{nahm}.

Each BPS-type equation can occur as the condition for the saturation of a bound 
on the tension of more than one worldvolume brane, since many of the latter are
related by duality. We shall therefore order the presentation of our results in
terms of the type of BPS equation, choosing the simplest case to derive the
bound that its solutions saturate. We shall also consider them roughly in order
of increasing complexity. We conclude with a brief discussion of some unresolved
puzzles.

\section{Abelian vortices}

We shall consider first a 0-brane soliton in the M-2-brane, arising from
intersections with other M-2-branes. This is associated with a scalar
central charge $Z$ in the M-2-brane's worldvolume supersymmetry algebra. It
should be possible to express $Z$ as an integral over the 2-dimensional
`worldspace' of a two-form constructed from the two worldvolume scalars defining
the 2-plane of the second M-2-brane. Let $U$ be a complex coordinate for this
2-plane. The (real) charge $Z$ must then take the form 
\be\label{abela}
Z = {i\over2}\int_{M2} dU \wedge d\bar U\, .
\ee
We shall now confirm the relevance of this charge by deriving a bound on the
energy of 0-branes within the M-2-brane.

The phase space Lagrangian density for the M-2-brane, in the M-theory vacuum and
omitting fermions, is \cite{M2ham}
\be
{\cal L} = P\cdot \dot X - s^a P\cdot \partial_a X - {1\over 2}v (P^2 + \det g), 
\ee
where all fields depend on the worldvolume coordinates $(t,\sigma^a)$
($a=1,2$). The Lagrange multiplier fields $s^a$ and $v$ impose the `worldspace'
diffeomorphism and hamiltonian constraints, respectively. In the `physical' or
`static' gauge ($X^0=t, X^a= \sigma^a$) the diffeomorphism constraint reduces to
\be
P_a = {\bf P}\cdot\partial_a {\bf X},
\ee
where ${\bf X}$ are the eight worldvolume scalars describing transverse
fluctuations and ${\bf P}$ are their conjugate momenta. In this gauge the
induced worldspace metric is
\be\label{worldspacemetric}
g_{ab} = \delta_{ab} + \partial_a {\bf X}\cdot \partial_b {\bf X}\, .
\ee
If we now restrict ourselves to static configurations, for which ${\bf P}={\bf
0}$, then $P_a=0$ too and so all components of the 11-momentum density $P$
vanish except $P^0={\cal E}$, the energy density. The hamiltonian constraint
now reduces to
\bea
{\cal E}^2 &=& \det g \nn
&=&1 + |\partial_1 {\bf X}|^2 + |\partial_2 {\bf X}|^2 +
|\partial_1 {\bf X}|^2 |\partial_2 {\bf X}|^2 - 
(\partial_1 {\bf X}\cdot \partial_2 {\bf X})^2.
\eea

Since we expect only two scalar fields to be of relevance
 we shall simplify our
task by setting to zero all but two of the eight scalars ${\bf X}$, in which
case\footnote { Retention of the others six scalars leads to additional
positive semidefinite terms which vanish when the six scalars are constants.}
\bea
{\cal E}^2 &=& 1 + |\vec \nabla X|^2 + |\vec \nabla Y|^2 + 
(\vec \nabla X \times \vec \nabla Y)^2 \nn
&=& (1 \pm \vec \nabla X \times \vec \nabla Y)^2 + 
|\vec \nabla X \mp \star\vec\nabla Y|^2,
\eea
where $X$ and $Y$ are the two scalars, and we use standard vector calculus for
$E^2$, with
\be 
\vec\nabla = (\partial_1,\partial_2) \qquad \star \vec \nabla =
(\partial_2,-\partial_1).
\ee
We deduce that
\be
{\cal E}-1 \ge | \vec \nabla X \times \vec \nabla Y |
\ee
with equality when
\be\label{cr}
\vec \nabla X = \pm \star\vec\nabla Y.
\ee

We may define the total energy relative to the worldvolume vacuum as
\be
{\it E} = \int_{M2} ({\cal E} -1)
\ee
The bound on ${\cal E}$ implies the bound\footnote { A related bound was 
discussed in \cite {GWG}.}
\be
{\it E} \ge |Z|
\ee
where $Z$ is the topological charge
\be
Z = \int_{M2} dX \wedge dY \, .
\ee
This bound will be saturated by solutions of (\ref{cr}) if, for these
solutions, the charge density is (positive or negative) semi-definite.
This condition is satisfied because for solutions of (\ref{cr}) the charge
density equals $\pm |\vec \nabla X|^2$. Thus, the total energy is bounded by
the magnitude of the charge $Z$ and is equal to $|Z|$ for solutions of
(\ref{cr}). 

The equations (\ref{cr}) are equivalent to the Cauchy-Riemann equations
for the complex function $U= X + iY$ of the complex variable $\sigma^1
+i\sigma^2$. In other words, the energy bound is saturated by holomorphic
functions $U$, and the energy is then the magnitude of a charge of the form    
(\ref{abela}), as claimed. Singularities of the holomorphic function $U$
represent `vortices' on the M-2-brane. These have been discussed in detail in  
in \cite{CM,GWG}; we shall call them `abelian vortices' for reasons that will
become apparent later. Essentially the same solution was used in \cite{HLW2} as
the M-5-brane worldvolume field configuration representing a 3-brane, which
therefore accounts for the form of the 3-brane charge given in (\ref{charges}).
The fact that the same solution serves in both contexts is to be expected
from the equivalence under spacetime duality of the intersecting brane
configurations associated with the M-2-brane 
vortex and the M-5-brane 3-brane. There
is, however, an additional feature of the energy bound in the M-5-brane context.
One must introduce constant worldvolume vector fields associated with the
3-brane worldvolume directions; one then finds that the saturation of the bound
requires the vanishing of the derivatives of all worldvolume fields in these
3-brane directions. This point will be illustrated later with the M-5-brane
string soliton so we pass over the details here. 

\section{D-brane solitons}

The phase-space form of the DBI worldvolume Lagrangian density, in the
D=10 spacetime vacuum and omitting fermions, is \cite{lind}
\bea
{\cal L} &=& P\cdot \dot X + E^a \dot V_a + V_t \partial_a E^a - s^a\big(
P\cdot \partial_a X + E^b F_{ab}\big)\nn
&& - {1\over 2}v \big[ P^2 + E^aE^b g_{ab} + \det (g+F)\big],
\eea
where $E^a$ is the worldspace electric field and $F_{ab}$ the worldspace
magnetic 2-form for the BI gauge potential $V= dt V_t + d\sigma^a V_a$. The
component $V_t$ imposes the Gauss law constraint on the electric field. The
Lagrange multipliers $s^a$ and $v$ impose the worldspace diffeomorphism and
hamiltonian constraints, respectively. 

In static gauge the $s$-constraint becomes
\be
P_a = -{\bf P}\cdot \partial_a {\bf X} - E^bF_{ab},
\ee
where ${\bf X}$ are the worldvolume scalars transverse to the D-brane and
${\bf P}$ their conjugate momenta. In addition, the worldspace metric $g_{ab}$
again reduces to the form given in (\ref{worldspacemetric}).
We shall now restrict to static configurations for which $\dot {\bf X} =0$ and
${\bf P}=0$. In this case the 10-vector $P$ is
\be
P = ({\cal E}, -E^bF_{ab}, {\bf 0})
\ee
where ${\cal E}$ is the energy density, and the hamiltonian constraint yields
\be
{\cal E}^2 = E^cE^dF_{ac}F_{bd} \delta^{ab} + E^aE^bg_{ab} + \det (g+F).
\ee
We now consider how this formula may be rewritten as a sum of squares, in
various special cases.

\subsection{BIons}

For purely electric configurations on the D-brane worldvolume we set $F_{ab}=0$
to get
\be\label{electrica}
{\cal E}^2 = E^aE^bg_{ab} + \det g.
\ee
We expect purely electric solutions to arise as endpoints of strings, the
string specifying a direction in the transverse space with coordinates ${\bf
X}$. We therefore set all but one of these transverse fluctuations to zero, in
which case 
\be
g_{ab} = \delta_{ab} + \partial_a X \partial_b X\, ,
\ee
where $X$ is the one non-zero scalar. Then $\det g = 1 + (\partial X)^2$
and we can rewrite (\ref{electrica}) as
\be\label{here}
{\cal E}^2 = (1\pm E^a\partial_a X)^2  + (E\mp\partial X)^2.
\ee
We deduce that 
\be
{\cal E}-1 \ge \big| E^a\partial_a X \big|,
\ee
with equality when
\be\label{BIBPS}
E_a = \pm \partial_a X.
\ee

The total energy ${\it E}$ relative to the worldvolume vacuum is
therefore subject to the bound
\be\label{BIbound}
{\it E} \ge |Z_{el}|
\ee
where $Z_{el}$ is the charge
\be\label{there}
Z_{el}= \int_{W} E^a\partial_a X 
\ee
where $W$ is the D-brane worldspace. The form of this charge is expected from
the fact that electric scalar charges in the worldvolume supersymmetry algebra 
are transverse 1-forms (arising from the reduction of the 10-momentum of the
N=1 D=10 supersymmetry algebra). The bound (\ref{BIbound}) on the energy is
saturated by  solutions of (\ref{BIBPS}) since the charge density is clearly
semi-definite for such solutions. 

Because of the Gauss law constraint on electric  field,
 solutions of (\ref{BIBPS}) 
correspond to solutions of $\nabla^2 X=0$, i.e. to harmonic functions on
worldspace. Isolated singularities of $X$ are the charged particle solutions
found in \cite{CM,GWG}. For D-P-branes with $P\ge 3$ the simplest solution is
\be
X = {q \over\Omega_{P-1} r^{P-2}}\, ,
\ee
corresponding to a charge $q$ at the origin, where $\Omega_{P}$ is the 
volume of the unit $P$-sphere.  
 Gauss's law allows us to write the
energy as an integral over a (hyper)sphere of radius $\epsilon$ surrounding
the charge. Since $X=X(\epsilon)$ is then constant over the integration region
we have
\bea
{\it E} &=&  \lim_{\epsilon \rightarrow 0}\big| X(\epsilon)
\int_{r=\epsilon} \vec {dS}\cdot \vec E \big| \nn
&=& q\; \lim_{\epsilon \rightarrow 0} X(\epsilon) 
\eea
As pointed out in \cite{CM,GWG}, the energy is infinite
since $X\rightarrow\infty$ as $\epsilon\rightarrow 0$, but the infinity has a
physical explanation as the energy of an infinite string of finite, and
constant, tension $q$. 

The D-string is a special case of particular interest. In
this case\footnote{To avoid possible confusion with the total energy, we remark
that the letter $E$ is reserved exclusively for the electric field in the
passage to follow.} 
\be
Z = \int_{D1} E X'
\ee
Gauss's law implies that the electic field $E$ is locally constant but it must
have a discontinuity at the endpoint of the F-string (F for `Fundamental')
because this endpoint carries electric charge. We may suppose that this charge
is at the origin $\sigma=0$ on the D-string and that $E=0$ when $\sigma < 0$.
If we further suppose that $X(0)=0$ then 
\be
Z= E \lim_{L\rightarrow\infty} X(L) 
\ee
where $E$ is the constant value of the electric field for $\sigma >0$. 
This is formally infinite because $X(L)$ grows linearly with $L$, but the
infinity again has a physical interpretation as the energy in an infinite string
of tension $E$. Since $E=dX$ we see that the D-string configuration with energy
$|Z|$ is
\be
X(\sigma)= \cases{0&$\sigma <0 $ \cr E\sigma & $\sigma >0 $}
\ee
so that a IIB F-string ending on a D-string produces, literally, a
kink in the latter. This point (in the context of the leading terms in the
expansion of the BI action) 
has recently been made independently 
in an
interesting paper \cite{mukhi} which we became aware of while writing up this
article. 

\subsection{BI Instanton}

Let us consider the case of the D-4-brane. The worldvolume superalgebra allows
scalar central charges in the ${\bf 1} \oplus {\bf 5}$ representations of the
$Spin(5)$ R-symmetry group, which we interpret as the transverse rotation
group. The scalars in the ${\bf 5}$ representation can be interpreted as the
endpoints of fundamental strings on the D-4-brane, which is the D-4-brane
subcase of the purely electric case considered above. The $Spin(5)$
singlet scalar is a magnetic charge, so we set the electric field to zero.
The magnetic charge on the D-4-brane corresponds to a spacetime configuration in
which a D-0-brane `intersects' a D-4-brane. Because it is a singlet its charge
cannot depend on any of the transverse scalars ${\bf X}$. We therefore set these
to zero. We then have
\bea\label{insta}
{\cal E}^2 &=& \det (\delta_{ab} +F_{ab}) \nn
&=& (1 \mp {1\over 4} \tr F\tilde F)^2 -   {1\over4}\tr (F\mp\tilde F)^2
\eea
where $\tilde F$ is the worldspace Hodge dual of $F$. The trace is over
the `worldspace' indices, i.e. $\tr F ^2 = F_{ab}F^{ba}$, but we can suppose it
to include a trace over $u(n)$ indices in the case of $n$ coincident 
D-4-branes. In either case we deduce that
\be
{\cal E} \ge 1 \mp {1\over4} \tr F\tilde F 
\ee
with equality when $F=\pm \tilde F$. The total energy ${\it E}$, relative to the 
worldvolume vacuum is therefore subject to the bound
\be
{\it E} \ge |Z|
\ee
where $Z$ is the topological charge
\be
Z= {1\over4}\int_{D4}  \tr F\tilde F\; ,
\ee
with equality when $F$ satisfies
\be
F= \pm \tilde F\; .
\ee
In the non-abelian case this is solved by (multi) instanton configurations. In
the abelian case, any solution must involve a singular BI gauge potential, but
the energy will remain finite as long as the charge is finite. 

\subsection{BI Dyons}

Consider now the D-3-brane. In this case, the endpoint of a (p,q) string
will appear in the worldvolume as a dyon. We should therefore allow for both
magnetic and electric fields, and one non-zero scalar.
Thus, we have
\be\label{monoa}
{\cal E}^2 =  E^aE^b(\delta_{ab} + \partial_a X\partial_b X) + \det
(\delta_{ab} + \partial_a X\partial_b X  + F_{ab}) + E^aE^b F_{ac}
F_{bd}\; \delta^{cd} 
\ee
Defining 
\be
B^a = {1\over 2}\varepsilon^{abc}F_{bc}
\ee
and expanding the $3\times 3$ determinant, we can rewrite this in standard vector
calculus notation as
\bea
{\cal E}^2 &=& 1 + |\vec\nabla X|^2 + |\vec E|^2 + |\vec B|^2 + (\vec E\cdot
\vec\nabla X)^2 + (\vec B\cdot \vec\nabla X)^2 + |\vec E\times \vec B|^2\nn
&=& (1+ \sin \vartheta \; \vec E \cdot \vec\nabla X + \cos\vartheta\; 
\vec B \cdot \vec\nabla X)^2 + |\vec E - \sin\vartheta\; \vec\nabla X|^2 +
|\vec B - \cos\vartheta\; \vec\nabla X|^2 \nn
&&+\, |\cos\vartheta\; \vec E\cdot \vec\nabla X - \sin\vartheta\; \vec B \cdot
\vec\nabla X|^2 + |\vec E\times \vec B|^2
\eea
where the second equality is valid for arbitrary angle $\vartheta$. 
We therefore deduce that
\be
{\cal E}^2 \ge (1+ 
\sin \vartheta \; \vec E \cdot \vec\nabla X + \cos\vartheta\; 
\vec B \cdot \vec\nabla X)^2
\ee
for arbitrary $\vartheta$. Taking the square root and then
integrating over the worldvolume of the D-3-brane we deduce that
the total energy,
relative to the D-3-brane vacuum, satisfies the bound
\be
{\it E} \ge \sqrt{ Z_{el}^2 + Z_{mag}^2}\; ,
\ee
where
\be
Z_{el} = \int_{D3} \vec E \cdot \vec\nabla X \qquad
Z_{mag} = \int_{D3} \vec B \cdot \vec\nabla X\, .
\ee
The bound is saturated when
\be\label{BPS}
\vec E = \sin\vartheta\; \vec\nabla X, \qquad 
\vec B = \cos\vartheta \vec\nabla X 
\ee
with
\be
\tan\vartheta = Z_{el}/Z_{mag}.
\ee
Since both $\vec E$ and $\vec B$ are divergence free (the latter as a 
consequence of the Bianchi identity), $X$ must be harmonic, i.e.
\be\label{harm}
\nabla^2 X =0
\ee
Given a harmonic function $X$, the electric and magnetic fields are then
determined by (\ref{BPS}). A BI dyon is then an isolated singularity of $X$. 
Again, all these formula have a natural non-abelian extension, but we shall
continue to assume a $U(1)$ group for ease of presentation.

For simplicity, let us
choose the solution for which
\be
\cos\vartheta X= {g\over 4 \pi r} 
\ee
where $r$ is the radial coordinate for $E^3$. Then the fact that $\vec E$ and
$\vec B$ are divergence free allows us to rewrite each charge as the
$\epsilon\rightarrow 0$ limit of an integral over a sphere of radius
$\epsilon$ centred on the origin. Since $X$ is constant over this sphere we
have
\be
Z_{el} = e\, \lim_{\epsilon\rightarrow 0} X(\epsilon) \qquad
Z_{mag} = g\, \lim_{\epsilon\rightarrow 0} X(\epsilon)
\ee
where we have set $e= g\tan\vartheta$. Thus
\be
{\it E} = \sqrt{ e^2 + g^2}\; \lim_{\epsilon\rightarrow 0} X(\epsilon)\, .
\ee
This is infinite since $X$ increases without bound as
$\epsilon\rightarrow\infty$, but the infinity has a physical interpretation as
the total energy of an infinite string of finite tension
\be
T = \sqrt{ e^2 + g^2}\, ,
\ee
as one expects for a $(p,q)$ string emanating from the D-3-brane.

\subsection{Non-abelian vortex}

The BPS equations for the BI magnetic monopole in the D-3-brane are just the
dimensional reduction of the self-duality equations for the BI instanton in the
D-4-brane. This follows from the fact that a D-0-brane in a D-4-brane is
T-dual to a D-string ending on a D-3-brane. If the starting point is a
single D-4-brane, with $U(1)$ gauge potential, then a further dimensional
reduction yields the abelian vortex equations discussed previously,
corresponding to the fact that the D-string ending on a D-3-brane is T-dual
to two D-2-branes intersecting at a point. This configuration is the obvious
reduction of the similar one involving two M-2-branes. However, we could
also start from a multi D-4-brane worldvolume with non-abelian gauge potential.
In this case T-duality takes the non-abelian instanton into a non-abelian
monopole. Further T-duality leads to what we may call a non-abelian vortex.
We may obtain the expression for the energy density by dimensional reduction of
(\ref{insta}). To this end, we define
\be
V_3 = X \qquad V_4 = Y
\ee
both in the adjoint representation of some non-abelian Lie algebra. Then
\be
{\cal E}^2 = \bigg[ 1 \pm  \tr \big( \star F [X,Y]- DX \times DY \big)\bigg]^2
+ \tr \big| DX \pm \star DY\big|^2 + \tr (\star F \mp [X,Y])^2
\ee 
we therefore deduce the bound
\be
{\cal E} -1  \ge \pm \big(  \star F [X,Y] - DX \times DY \big)
\ee
which is saturated when
\be\label{hitchsys}
DX =\mp \star DY \qquad \star F = \pm [X,Y]
\ee
which are the equations studied by Hitchin \cite{hit}.
The total energy is therefore subject to the bound
\be
{\it E} \ge |Z|
\ee
where $Z$ is the topological charge
\be
Z= \int_{D2} \tr\big( {1\over 2}DX^2 +{1\over 2} DY^2 + [X,Y]^2\big)
\ee

\subsection{Nahm's Equations}

A further T-duality of the two intersecting D-2-branes leads us back to a
D-string ending on a D-3-brane, but we now find BPS-type equations
for the {\sl string's} worldvolume field. These are just the dimensionally
reduced version of Hitchin equations (\ref{hitchsys}). Equivalently, they are
the self-duality equations for non-abelian $F$ on the D-4-brane, dimensionally
reduced in three orthogonal directions. Let $X_i = V_i,\; i=1,2,3$ (the
components of the Lie-algebra valued gauge potential in the three directions)
and define $D$ to be the covariant derivative in the 4-direction. Then we
arrive at  
\be\label{Nahm}
D X_i =\pm {1\over 2}\varepsilon_{ijk}[X_j,X_k],
\ee
which are Nahm's equations. Solutions of these equations are associated
with an energy given by
\be\label{blip}
{\it E} = \frac 12 \big|\int_{D1} \varepsilon _{ijk}
tr(DX_i[X_j,X_k])\big|\; .
\ee
and represent the
intersection of multiple D-strings with D-3-branes. Since the 
D-strings end on the D-3-branes we must impose suitable boundary
conditions on the worldline fields. 
In \cite{Diaconescu} it was argued in
the context of the leading order terms in the BI action that the
appropriate supersymmetric boundary conditions are precisely those that lead
to the Nahm description of the moduli spaces of $SU(k)$ monopoles,
where $k$ is the number of D-3-branes.
That the arguments in \cite{Diaconescu}
are valid for the full BI gauge theory is supported by our results.

A further dimensional reduction leads to BI
quantum mechanics for $U(n)$ matrices describing the dynamics of
$n$ D-0-branes. The
energy bound becomes 
\begin{equation}\label{bd}
E\ge \frac 12|\varepsilon_{ijkl}tr(X_i X_j X_k X_l)|
\end{equation}
where here $X_i=V_i$, $i=1,\dots,4$, and the bound is saturated for matrices satisfying 
\begin{equation}
[X_i,X_j]=\frac 12 \varepsilon_{ijkl}[X_k,X_l].
\end{equation}
Using the cyclic property of the trace we see that the right hand
side of (\ref{bd}) vanishes for finite dimensional matrices.
Note that for infinite dimensional matrices such charges
correspond to longitudinal fivebranes in the matrix theory approach
to M-theory \cite{ORT,BSS}.

\section{String-in-fivebrane}

The phase space Lagrangian density for the M-5-brane, in the M-theory vacuum
and omitting fermions, is \cite{bergtown}
\bea
{\cal L} &=& P\cdot \dot X + \Pi^{ab} \dot A_{ab} + \lambda_a \partial_b \Pi^{ab}
- s^a\big( P\cdot \partial_a X -V_a \big) \nn
&& + \sigma_{ab}(\Pi^{ab} + {1\over4}\tilde{\cal H}^{ab}) - {1\over 2}v 
\big[ (P- g^{ab}V_a\partial_b X)^2 + \det (g+ \tilde H) \big]
\eea
where
\bea
\tilde {\cal H}^{ab} &=& {1\over6} \varepsilon^{abcde} H_{cde}\nn
\tilde H_{ab} &=& {1\over \sqrt{\det g}} g_{ac} g_{bd} \tilde {\cal H}^{cd}\nn
V_f &=& {1\over 24} \varepsilon^{abcde} H_{abc}H_{def} 
\eea
with $\varepsilon$ the invariant worldspace tensor density (such
that $\varepsilon^{12345}=1$). Note that $\lambda_a$ imposes the Gauss law
constraint on the electric 2-form $\Pi$. This becomes equivalent to the Bianchi
identity $dH=0$ upon using the constraint imposed by the Lagrange multiplier
$\sigma_{ab}$. 

In static gauge, and restricting to static configurations, the $s$-equation
implies $P_a=V_a$, so the hamiltonian constraint becomes
\be\label{foura}
{\cal E}^2 = V_aV_b m^{ab} + \det (g_{ab} + \tilde H_{ab})
\ee
where
\be
m^{ab} = g^{aa'}g^{bb'} \big[ \partial_{a'} {\bf X} \cdot \partial_{b'}{\bf X}
+ (\partial_{a'}{\bf X}\cdot \partial_{c}{\bf X})\delta^{cd}
(\partial_{b'}{\bf X}\cdot \partial_{d}{\bf X})\big]\; .
\ee
The expansion of the determinant leads to terms quartic in $\tilde H$, but 
the identity
\be
\det (\tilde H ) \equiv  V_aV_b g^{ab}
\ee
allows these terms to be combined with the other terms quadratic in $V$,
leading to the result
\be\label{fourb}
{\cal E}^2 =  \det g  + {1\over2} \tilde {\cal H}^{ac}\tilde {\cal
H}^{bd}g_{ab}g_{cd} + V_aV_b \delta^{ab}
\ee

We shall now set all but one of the transverse scalars to zero, in which case
\be
{\cal E}^2 = 1 + (\partial X)^2 + {1\over2} |\tilde {\cal H}|^2 + 
|\tilde {\cal H}\cdot \partial X|^2 + |V|^2 
\ee
where $X$ is the one non-zero scalar field and
\bea
|\tilde {\cal H}|^2 &=& {\cal H}^{ab} {\cal H}^{cd} \delta_{ac} \delta_{bd}\nn
|\tilde {\cal H}\cdot \partial X|^2 &=& 
\tilde {\cal H}^{ab}\tilde {\cal H}^{cd}\partial_b X\partial_d X\delta_{ac} \nn
|V|^2 &=& V_aV_b \delta^{ab}
\eea
We can rewrite this as
\bea
{\cal E}^2 &=& \big|\zeta^a \pm \tilde {\cal H}^{ab}\partial_b X\big| ^2 
+ 2\bigg|\partial_{[a} X \zeta_{b]} \pm {1\over2} \delta_{ac}\delta_{bd} \tilde
{\cal H}^{cd}\bigg|^2 \nn
&& +\, (\zeta^a\partial_a X)^2 + |V|^2 
\eea
where $\zeta$ is a unit length worldspace 5-vector, i.e.
\be
\zeta^a \zeta^b\delta_{ab} =1\, .
\ee
We may choose $\zeta^5=1$ and $\zeta^{\hat a}=0,\; (\hat a =1,2,3,4)$, in which
case we deduce the inequality
\be
{\cal E} -1 \ge \pm {1\over 6}\varepsilon^{\hat a\hat b\hat c\hat d}H_{\hat
a\hat b\hat c} \partial_{\hat d}X
\ee
with equality when
\be\label{stringsola}
\partial_5 X=0 \qquad H_{5\hat a \hat b} =0
\ee
and
\be\label{stringsolb}
H = \pm \star dX  
\ee
where, in the last equation, $H$ is restricted to the 4-dimensional subspace of 
worldspace orthogonal to $\zeta$, which we shall call $w_4$, and $\star$ is the
Hodge dual of $w_4$. 

Imposing periodic boundary conditions in the 5-direction so that the vector field
$\zeta$ has orbits of length $L$, we see that the total energy satisfies the
bound 
\be
{\it E} \ge L\times |Z|
\ee
where $Z$ is the topological charge
\be
Z= \int_{w_4} H\wedge dX\, .
\ee
The tension $T={\it E}/L$ is therefore bounded by $Z$, as claimed in the
introduction, with equality for configurations satisfying (\ref{stringsolb}).
Because $H$ is closed, this equation implies that $X$ is harmonic.
Singularities of $X$ are the strings found in \cite{HLW1}.
The simplest solution is obtained by choosing a single isolated point
singularity at the origin. In this case the energy integral can be rewritten as
the small radius limit of a surface integral over a 3-sphere surrounding the
origin. Since $X$ is constant on this surface we deduce that the string tension
is given by
\be
T = \mu \lim_{\epsilon\rightarrow 0}  X(\epsilon) 
\ee
where
\be
\mu = \int_{S^3} H
\ee
is the `naive' string charge defined by the 3-form flux through a 3-sphere
surronding the string in the 5-brane. The tension of this string is
formally infinite since $X(\epsilon)$ increases without bound as
$\epsilon\rightarrow 0$, but this is precisely what is expected if the string
is the boundary of a semi-infinite membrane.

\section{Discussion}

There is a class of brane intersections for which no worldvolume soliton is
known. This class includes the case of two M-5-branes intersecting on a
string, two D-4-branes intersecting on a point, and a D-0-brane in
a D-8-brane. In the latter case, the 0-brane charge is expected to be
\be
Z= \int_{D8} F\wedge F\wedge F\wedge F
\ee 
The charge in the two M5-brane case should be similar, and the other cases,
which are related by duality, are presumably obtained by dimensional reduction.
By analogy with the instanton case, we would expect a purely magnetic solution
with all scalars constant. For such configurations the energy density is given
by ${\cal E}^2 =\det(\delta +F)$ which can be rewritten as
\be\label{inste}
{\cal E}^2 = (1 \pm \star F^4)^2 +{1\over2} (F\mp \star F^3)^2 + {1\over 2.
4!}(F^2 \pm \star F^2)^2
\ee
where
\bea\label{instb}
\star F^4 &=& {1\over 2^4 4!} \varepsilon^{ijklmnpq}F_{ij}F_{kl}F_{mn}F_{pq}\nn
(\star F^3)^{ij} &=& {1\over 2^3 3!} \varepsilon^{ijklmnpq}F_{kl}F_{mn}F_{pq}\nn
(\star F^2)^{ijkl} &=& {1\over 8} \varepsilon^{ijklmnpq}F_{mn}F_{pq}\nn
(F^2)_{ijkl} &=& 3F_{[ij}F_{kl]}\; .
\eea
We conclude that
\be
{\it E}\ge |Z|
\ee
but the bound cannot be saturated because the equations
\be
F= \pm \star F^3 \qquad F^2 = \mp \star F^2
\ee
have no simultaneous solutions\footnote{If the signs were different a solution 
of the form considered in \cite{wash} would be possible.}. We suspect that a
resolution of this problem will involve the $VF^4$ Chern-Simons term for the BI
field that is required in a massive IIA background \cite{green}, but we leave
this to future investigations. 

We should not conclude without a comment on the remarkable fact that the
BPS-type equations for ( supersymmetric) worldvolume solitons are 
{\it linear}, despite the non-linearity of the action. The seems to be 
a special feature of M-brane and D-brane worldvolume actions.

\vskip 0.5cm
\noindent
{\bf Acknowledgements}: We thank David Mateos for helpful discussions.
PKT gratefully acknowledges the support of the Iberdrola 
{\it Profesor Visitante} program. JPG thanks the members of the Faculty of
Physics at the University of Barcelona for hospitality
and the EPSRC for an Advanced Fellowship. 
This work has been
partially supported by AEN95-0590 (CICYT), GRQ93-1047 (CIRIT) and by the
Commission of European Communities CHRX93-0362 (04).


\begin{thebibliography}{99}

\bibitem{polch}
J. Polchinski, {\sl Dirichlet branes and Ramond-Ramond charges}, Phys. Rev.
Lett. {\bf 75} (1995) 184.

\bibitem{Strom}
A. Strominger, {\sl Open p-branes}, Phys. Lett. {\bf 383B}, (1996) 44.

\bibitem{PKT}
P.K. Townsend, {\sl D-branes from M-branes}, Phys. Lett. {\bf 373B}, (1996) 68;
{\sl Brane Surgery}, {\it Nucl. Phys.} (Proc. Suppl.) {\bf B58} (1997)
163, hep-th/9609217.

\bibitem{paptown} 
G. Papadopoulos and P.K. Townsend, {\sl Intersecting
M-branes}, Phys. Lett. {\bf 380B} (1996) 273, hep-th/9603087.

\bibitem{CM}
C.G. Callan and J.M. Maldacena, {\it Brane dynamics from the Born-Infeld
action}, hep-th/9708147.



\bibitem{HLW1}
P.S. Howe, N.D. Lambert and P.C. West, {\it The self-dual string soliton},
hep-th/97009014.

\bibitem{GWG}
G.W. Gibbons, {\it Born-Infeld particles and Dirichlet p-branes},
hep-th/9709027.

\bibitem{HLW2}
P.S. Howe, N.D. Lambert and P.C. West, {\it The threebrane soliton of the
M-fivebrane}, hep-th/9710033.

\bibitem{BGT}
E. Bergshoeff, J. Gomis and P.K. Townsend, {\sl M-brane intersections from
worldvolume supersymmetry algebras}, hep-th/9711043.

\bibitem{DVV}
R. Dijkgraaf, E. Verlinde and H. Verlinde, {\it BPS quantisation of the five-brane},
Nucl. Phys. {\bf B486} (1997) 89.

\bibitem{azc}
J.A. de Azcarraga, J.P. Gauntlett, J.M. Izquierdo 
and P.K. Townsend , Phys. Rev. Lett. {\bf 63} (1989) 2443.

\bibitem{sorokin}
D. Sorokin and P.K. Townsend , {\sl M Theory Superalgebra from the M-5-brane},
hep-th/9708003.




\bibitem{Bog}
E.B. Bogomol'nyi, Sov. J. Nucl. Phys. {\bf 24} (1976) 449.




\bibitem{peet}
S. Lee, A. Peet, L. Thorlacius, {\sl Brane-waves and strings}, hep-th/9710097.


\bibitem{Nakamura}
A. Nakamura and K. Shiraishi, {\sl Born-Infeld monopoles and instantons},
Hadronic Journal, {\bf 14}, (1991) 369.

\bibitem{Tseytlin}
A.A. Tseytlin, {\sl On non-abelian generalization of Born-Infeld action 
in string theory}, Nucl. Phys {\bf B501} (1997) 401.


\bibitem{nahm}
W. Nahm {\sl A Simple Formalism for the BPS Monopole}, Phys. Letts. {\bf 90B}
(1980) 413.

\bibitem{M2ham}
E. Bergshoeff, E. Sezgin and Y. Tanii, {\sl Hamiltonian formulation of the
supermembrane}, Nucl. Phys. {\bf B298} (1988) 187.

\bibitem{lind}
U. Lindstr{\"o}m and R. von Unge, {\it A picture of D-branes at
strong coupling}, Phys. Lett. {\bf 403B} (1997) 233; R. Kallosh, {\sl Covariant
quantization of D-branes}, hep-th/9705056; K. Kamimura and M. Hatsuda, {\sl
Canonical formalism for IIB D-branes}, to appear. 

\bibitem{mukhi}
K. Dasgupta and S. Mukhi, {\sl BPS nature of 3-string junctions},
hep-th/9711094.


\bibitem{hit}
N.J. Hitchin, {\sl The self-duality equations on a Riemann surface},
Proc. Lond. Math. Soc. {\bf 55} (1987) 59.

\bibitem{Diaconescu}
D-E. Diaconescu, {\sl D-branes Monoples and Nahm Equations}, 
Nucl. Phys. {\bf 503B} (1997) 220.

\bibitem{ORT}
O.J. Ganor, S. Ramgoolam and W. Taylor IV, {\sl Branes, fluxes and duality 
in (M)atrix theory}, Nucl.Phys. {\bf B492} (1997) 191.

\bibitem{BSS}
T. Banks, N. Seiberg and S. Shenker, {\sl Branes from matrices},
Nucl.Phys. {\bf B490} (1997) 91.

\bibitem{bergtown}
E. Bergshoeff and P.K. Townsend, to appear.

\bibitem{wash}
W. Taylor, {\sl Adhering 0-branes to 6-branes and 8-branes},
hep-th/9705116.

\bibitem{green}
M.B. Green, C.M. Hull and P.K. Townsend, {\sl D-brane WZ terms, T-duality and
the cosmological constant}, Phys. Lett. {\bf 382B} (1996) 65.





\end{thebibliography}
\end{document}